\documentclass[11pt]{article}

\usepackage[T1]{fontenc}

\usepackage[utf8]{inputenc}

\usepackage{lmodern}

\usepackage{amsmath,amssymb,graphicx,booktabs,natbib, url, hyperref}

\usepackage{tabularx}

\title{A comparison of human and LLM-simulated participants in a writing style task}

\author{Felix Gr\"{o}ner and Erin K. Chiou\\
\small Arizona State University, Mesa, Arizona, USA
}

\date{}

\begin{document}

\maketitle

\begin{abstract}

Because large language models (LLMs) can produce natural language that is sometimes indistinguishable from texts produced by people, some researchers are starting to consider replacing human participants with LLM simulations. In this study, we test the extent to which the findings of a simulation with an LLM prompted to act as a synthetic participant match those obtained from 30 human participants. In our experiments, we evaluated how well writing style preference inference algorithms adapted to a participant over repeated interactions, compared to a baseline. We discover hints of bias and a lack of depth in GPT-4o's text generation and judgement that prevent it from accurately simulating people's behavior. Our results also hint at human biases that highlight the importance of considering human factors in the evaluation of systems that depend on human-automation interaction. Rather than treating these discrepancies as evidence for or against the validity of LLM-simulated participants, we present this study as a case analysis of methodological and design challenges.

\end{abstract}

\section{Introduction}\label{sec:introduction}

Large language models (LLMs) can produce texts that are superficially indistinguishable from those produced by people. This raises the question whether their reasoning, decision-making, and behavior are sufficiently similar that simulations with LLM-simulated participants lead to the same findings and conclusions as experiments with human participants. If this were the case, LLM-powered simulations could drastically lower the cost and increase the speed of research. Some research groups have started to empirically test novel algorithms and systems in simulations in which LLMs are prompted to act like people interacting with the system. However, we are concerned to what extent such simulations and the findings derived from them generalize to a population of real people.

To investigate this issue, we compared the results of a simulation and an experiment with human participants using a writing style preference inference paradigm. This use-case is particularly interesting for our research question because it places higher requirements on LLMs to faithfully replicate people's behavior.

We found that our synthetic participants could not sufficiently simulate people's writing and editing behaviors. In our setup, GPT-4o failed to replicate human biases and introduced its own biases, leading to a discrepancy between the findings from the simulation and the experiment with human participants. We present our findings not as a general rejection of LLM simulations as a research method but as a cautionary tale of experimental design issue surrounding them.

\section{Related work}\label{sec:related-work}

\subsection{LLMs simulating people's behavior}\label{sec:llm-simulating-people}

Several studies have tested the viability of LLM-powered simulations for replicating human participants. A common question is how closely the responses and behaviors of LLMs resemble those of people. A review by \citet{lu_llms_2024} provides several examples of experiments that were successfully replicated by LLM agents. LLMs were also able to match people's responses in economic games \citep{horton_large_2023,xie_can_2024}, movie selections \citep{wang_user_2025}, moral judgements \citep{dillion_can_2023}, and political settings \citep{argyle_out_2023}. A common finding is that simulations come close, but are not perfect. They might correctly predict the effect direction but not its strength \citep{manning_automated_2024} or have impressive but insufficient accuracy \citep{aher_using_2023,cui_can_2025,petrov_limited_2024}. Many studies have also reported significant differences between models \citep{aher_using_2023,petrov_limited_2024,xie_can_2024}. 

Most of these studies aimed to replicate general, not individual, human behavior on a rather superficial level. \citet{petrov_limited_2024} showed that GPT-4 fails to replicate a specific individual's responses to psychometric questionnaires when provided with the person's extensive biographic information. Simulated users may not reflect real human variability, biases, motivations, or constraints \citep{cheng_compost_2023}. In fact, they are likely to introduce their own biases and constraints, such as sensitivity to changes in prompts, hyper accuracy, and biases stemming from the training data as well as the network architecture \citep{aher_using_2023,anthis_llm_2025,wang_what_2025,yang_llm_2024,ye_justice_2024,zheng_judging_2023}.

These insights from related work lead us to hypothesize that current state-of-the-art LLMs cannot accurately simulate human behavior in all domains and settings. We aim to establish whether this is the case for writing style preferences in article summarization tasks.

\subsection{Preference inference for LLM personalization}\label{sec:preference-inference}

The purpose of personalization is to provide a better user experience by producing responses tailored to a specific individual rather than generic ones. In other words, the desired output for the same prompt may differ amongst individuals. Thus, a one-size-fits-all output introduces inefficiencies by requiring some people to spend additional editing effort until the generic output matches the personally desired output. Several approaches have been attempted to personalize the output of LLMs, which can be roughly divided into fine-tuning \citep{stiennon_learning_2020,tan_democratizing_2025,zhuang_hydra_2024} and in-context prompting \citep{dong_can_2024,lin_unlocking_2023,salemi_lamp_2024,shen_pmg_2024}. More rigorously developed taxonomies can be found in dedicated reviews \citep{liu_survey_2025,zhao_llms_2025,jiang_survey_2024,zhang_personalization_2024}.

To evaluate how well a system can infer people’s implicit writing styles, \citet{gao_aligning_2024} introduced the PRELUDE (PREference Learning from User's Direct Edits) framework in which an LLM observes an individual's revisions across multiple writing tasks (e.g., summarizing articles, composing emails) and adapts future outputs with the goal to minimize the editing effort. They compare different algorithms and find that a combination of retrieving context-specific examples and inferring preferences in a separate step before generating the summary resulted in the lowest editing effort \citep{gao_aligning_2024}. 

This setup has been criticized for merely requiring participants to edit summaries rather than write them from scratch \citep{aroca-ouellette_predict_2024}. The core argument is that pre-existing summaries will prime people and heavily influence their stylistic choices. This form of the anchoring bias would lead individuals to come to an acceptable outcome rather than the best outcome with the least amount of effort. It seems plausible that this effect would be stronger for human editors than for simulated editors. However, requiring people to write summaries from scratch to optimize the inference process is likely to worsen the user experience.

\section{Experiment 1: Simulation}\label{sec:experiment1}

To identify possible shortcomings of LLM simulations, and gather evidence for or against the validity of such research methods, we conducted two similar experiments with synthetic participants (i.e., an LLM prompted to simulate a user) and real participants. This study is heavily inspired by the work of Gao et al., although it is not a faithful replication. While we used the performance of different algorithms as the dependent variable in both experiments, this comparison was not our main focus. Instead, our research question is to what extent the findings of the two experiments match and in which details they differ. Therefore, we used simplified versions of the PRELUDE framework and preference inference algorithms.

\subsection{Methods}\label{sec:methods2}

For the first experiment, we set up a simulation with synthetic participants to compare different personalization algorithms. The three core steps repeated in the simulation were as follows: 

\begin{enumerate}

\item The system generates a summary for an article.

\item The user checks whether the summary is appropriate for their target audience.

\item If the user decides that the summary must be edited, they generate a revised version.

\item Depending on the algorithm, the pair of original and edited summaries is used to infer preferences which are taken into consideration for future summary generations.

\end{enumerate}

We collected 20 articles from Wikipedia and edited them to be of approximately equal length (ranging from 136 to 293 words; full list is provided in Appendix~\ref{app:a}). The order of articles was randomized for each simulation run. We used OpenAI's \texttt{gpt-4o-2024-08-06} \citep{openai_hello_2024} with a temperature of 1.0 for all generations. The simulation has two components: first, the system that generates the summaries, observes the changes made by the user, infers the writing style preferences, and then takes those into consideration for the next summary; and second, the synthetic user who receives a summary and edits it if it does not match their preferences.

For each article, an ad-hoc summary was generated using one of three algorithms, which are described in detail below. The summary was then passed on to a simulated user who was tasked with ensuring that it is appropriate for one of four target audiences, which are also described below. We repeated this to have three runs across the 20 articles for each of the four target audiences for each of the three algorithms. The personas and prompt templates are heavily inspired by those originally used by \citep{gao_aligning_2024} as well as revisions in \citep{aroca-ouellette_predict_2024}, and can be found in Appendices \ref{app:b} and \ref{app:c}.

\subsubsection{Synthetic participants and target audiences}\label{sec:synthetic-participants}

We followed Gao et al.'s two-step design for the editing agents. In a first step, the LLM is prompted to decide whether a given original summary is appropriate for its designated target audience. If it responds with ``no,'' then it is prompted to edit the summary to make it appropriate.

Each synthetic user is assigned to one target audience and does not change their style preferences over the course of the 20 articles. This is a simplification of the PRELUDE setup in Gao et al. but is sufficient for our purposes. The four target audiences ``young children watching TV,'' ``high-school students in class,'' ``managers and CEOs,'' and ``young adults on social media'' were designed and accompanied by instructions to induce between-group variance in appropriate writing styles. 

\subsubsection{Summary-generating system}\label{sec:summary-generating-system}

\begin{figure}[t]

\centering

\includegraphics[width=0.95\linewidth]{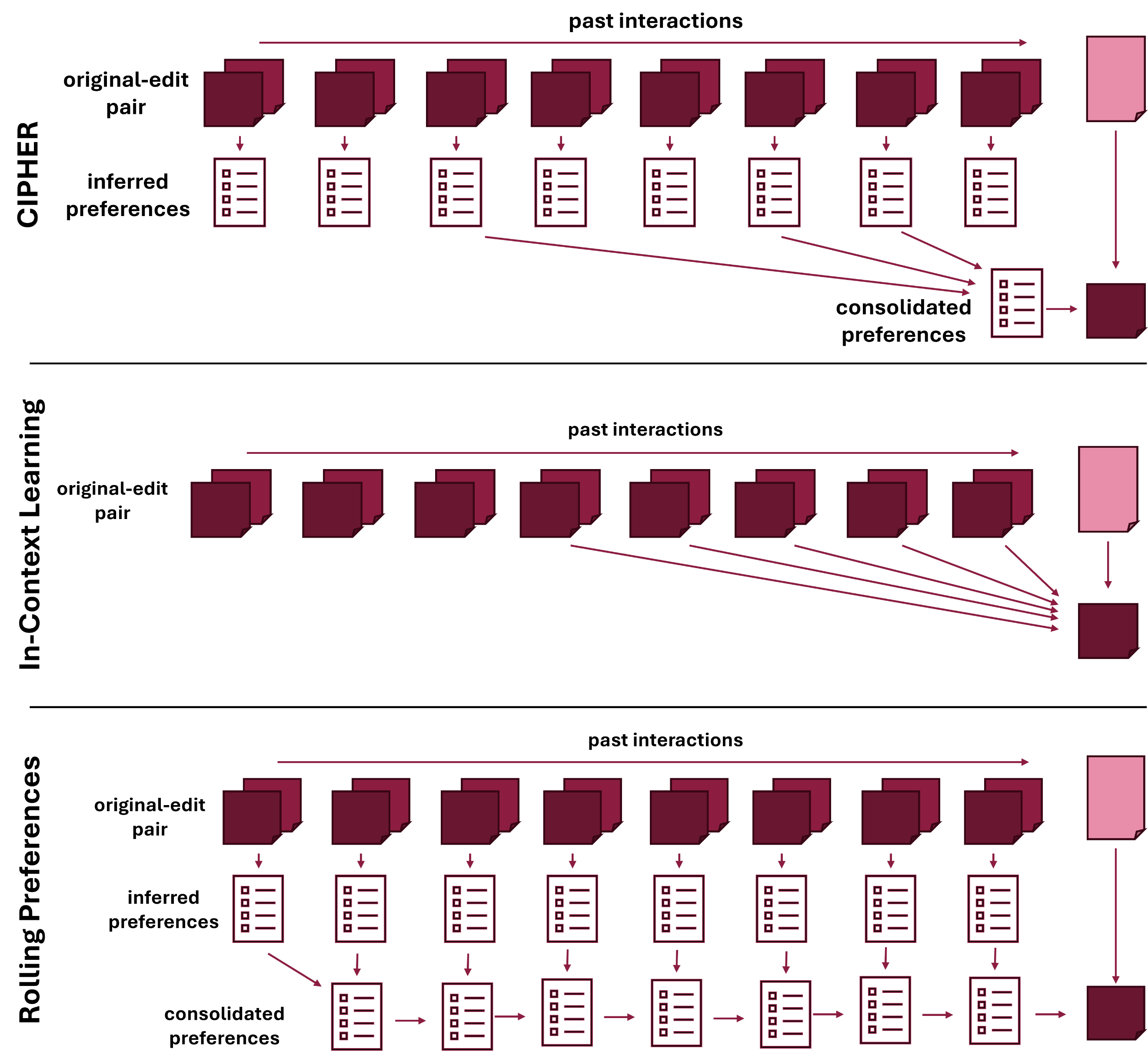}

\caption{Comparison of preference inference algorithms.}

\label{fig:pref-algorithms}

\end{figure}

We compared two preference inference algorithms (in-context learning and rolling preferences) and a baseline (no personalization). We curated this selection to include conditions for which prior work found significant performance differences \citep{aroca-ouellette_predict_2024,gao_aligning_2024}. The no-personalization baseline simply produces a summary without inferring or considering the user's preferences. The two preference inference algorithms are illustrated in Figure~\ref{fig:pref-algorithms}. For in-context learning (ICL), the five previous pairs of original and edited summaries were inserted into the prompt as examples to guide the generation of the next summary. This format leverages LLMs' ability to infer patterns and rules from examples \citep{brown_language_2020}. In this algorithm, the preferences are never explicitly stated.

Rolling preferences (RP) is prompted in three steps. It keeps a list of the inferred preferences for a specific user (initially empty) and considers them when generating the summary. It is then prompted with the original and edited versions of a summary to infer the preferences from the edits made by the user in this specific case. Finally, the general preferences of this user are updated by consolidating them with the newly inferred preferences. 

Note that these two inference algorithms are similar to those presented in prior works \citep{aroca-ouellette_predict_2024,gao_aligning_2024} but do not reflect the state-of-the-art. For simplification, we omitted the context-relevant retrieval used in these studies to adapt to changing preferences.

\subsubsection{Metrics}\label{sec:metrics}

We used two metrics to evaluate the different algorithms: normalized Levenshtein distance and BERTScore. The two reference texts were always the original and edited summaries. Levenshtein distance is defined as the minimal number of operations (insertions, deletions, and substitutions) required to change one text to another \citep{levenshtein_binary_1966}. We calculate this on the token level using \texttt{cl100k\_base} encoding. This can be interpreted as a proxy for the user's editing effort to convert the initial summary into an appropriate style. BERTScore is a measure of semantic similarity between two texts \citep{zhang_bertscore_2020}. In this study, we used the F1 score produced by the \texttt{roberta-large} model. This can be interpreted as reflecting the match of information included in the summaries. Using these metrics allowed us to evaluate the hundreds of summary pairs from this experiment more quickly, cheaply, consistently, and objectively than manual scoring.

\subsection{Results}\label{sec:results1}

\begin{figure}[t]

\centering

\includegraphics[width=0.95\linewidth]{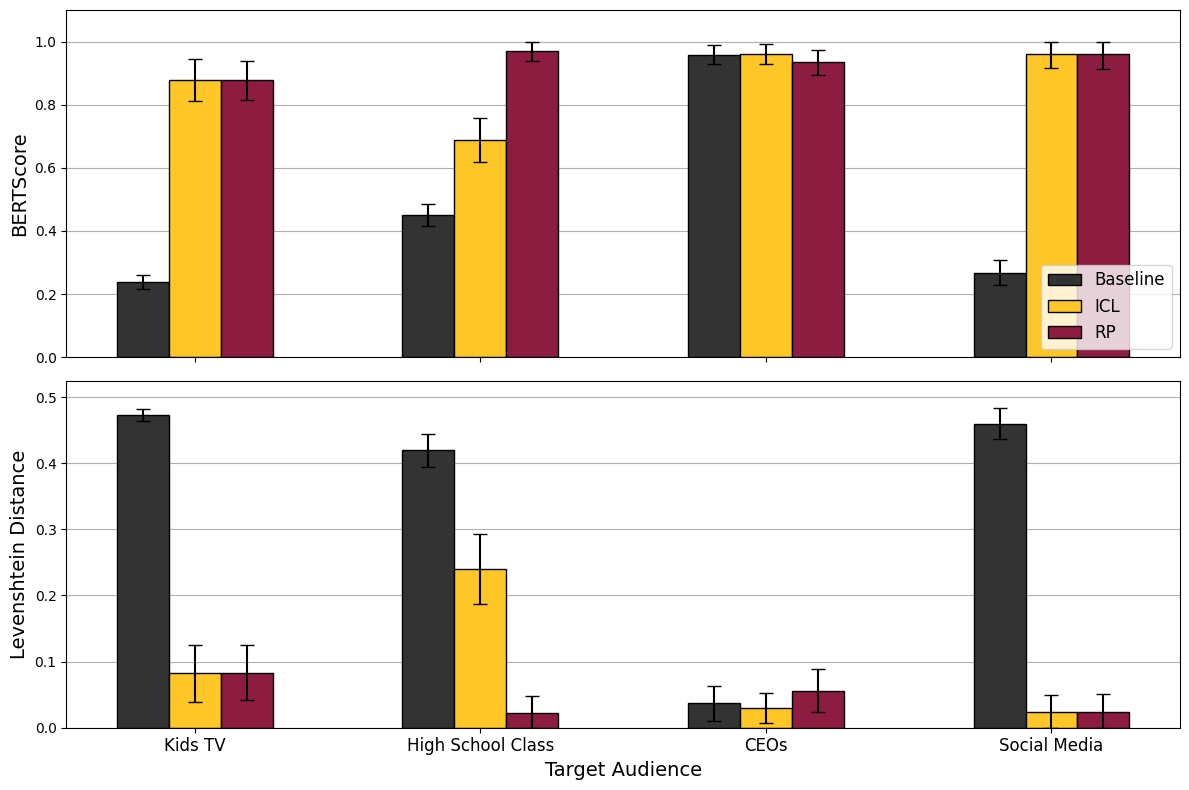}

\caption{Summary evaluations for different preference inference algorithms across all articles. \textit{Similarity scores were calculated by comparing the summary originally generated by the algorithms to the edited version generated by the simulated participant, and averaged across all iterations. Error bars indicate 95\% confidence intervals. ICL = in-context learning, RP = rolling preferences.}
}

\label{fig:summary-evals}

\end{figure}

We report our findings mostly with regard to BERTScore, because a\linebreak repeated-measures correlation analysis \citep{vallat_pingouin_2018}, considering the LLM in each condition as an individual subject, showed a strong association between BERTScore and Levenshtein distance in our generated dataset, $r(718)=-0.99$, $p<.001$. Figure~\ref{fig:summary-evals} shows the average evaluation scores for the different preference inference algorithms averaged for each target audience. Interestingly, our results do not show a consistent pattern across all target audiences. In each condition, the two preference inference algorithms performed roughly equally, with the notable exception of ``high school class,'' where ICL underperformed. Furthermore, the RP algorithm performed better than the baseline for all target audiences except for CEOs. Here, the baseline performed approximately on par with the preference inference algorithms. This indicates that GPT-4o's default summaries are close to what GPT-4o deems appropriate for CEOs.

\begin{figure}[t]

\centering

\includegraphics[width=0.95\linewidth]{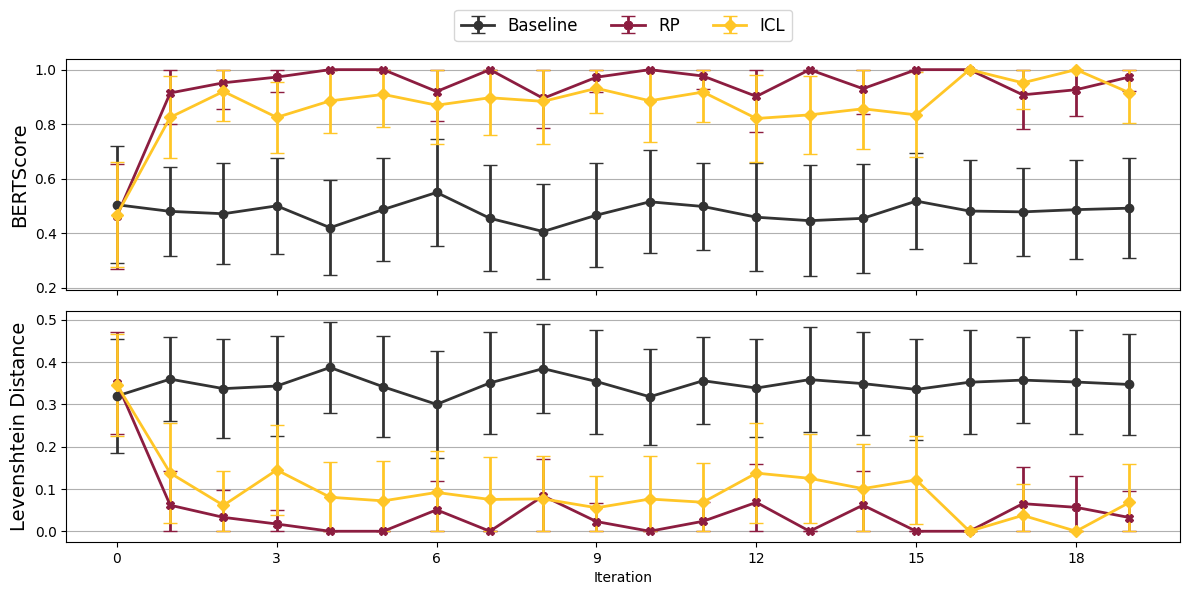}

\caption{Development of summary evaluation metrics over repeated interactions in the simulation. \textit{Similarity scores were calculated by comparing the summary originally generated by the algorithms to the edited version generated by the simulated participant. Error bars indicate 95\% confidence intervals. ICL = in-context learning, RP = rolling preferences.}}

\label{fig:sim-metrics}

\end{figure}

Figure~\ref{fig:sim-metrics} shows the development of the summary evaluation metrics over repeated interactions. Two-sided pairwise t-tests of BERTScores at different iterations showed that the preference inference algorithms outperformed the baseline after just one iteration, $t(11) = -4.71$, $p < .001$, for RP, and $t(11) = -3.17$, $p < .001$ for ICL. However, their scores stagnated for the remainder of the simulation (comparing the second to the final iteration), $t(11) = -0.87$, $p = .403$, for RP, and $t(11) = -0.83$, $p = .425$ for ICL. For ICL, one could have expected a more gradual improvement as more examples are added to the limit of five. Note that these results include cases in which the simulated user decided that the originally generated summary already was appropriate and made no further changes, resulting in a BERTScore of 1 and Levenshtein distance of 0. The results without this step (i.e., asking the simulated user to edit without checking whether the summary was already appropriate) are provided in Appendix~\ref{app:d}. They show over-eagerness to rewrite, resulting in worse metrics. An illustrative example of this is provided in Appendix~\ref{app:e}.

\section{Experiment 2: Human participants}\label{sec:experiment2}

We replicated the setup of the first experiment with human participants instead of simulated users. We used the insights from the simulation to reduce the scope (number of conditions and workload for participants) of this experiment.

\subsection{Methods}\label{sec:methods1}

\subsubsection{Procedure}\label{sec:procedure}

Participants were tasked with reading 10 ad-hoc generated summaries and revising them if necessary to make them appropriate for a specific target audience. We opted for this reduced number of summaries after observing the early stagnation in the simulations. These articles were drawn from the original set of articles (indicated in Appendix~\ref{app:a}) and made available to participants even though they were not necessary to complete the task. 

Participants accessed the online study through a custom-built platform. We did not monitor or restrict the environment in which participants completed the study or which devices they used. On each page, participants were presented with the original article, the ad-hoc generated summary, and a reminder of their target audience. Participants rated the AI summary's appropriateness for their assigned audience on a 0--100 scale, then revised the summary in a pre-filled text box to better suit their audience's needs. Instructions given to participants are provided in Appendix~\ref{app:h} and a screenshot of participants' view during the experiment is shown in Figure~\ref{fig:screenshot} of Appendix~\ref{app:i}.

\subsubsection{Participants and conditions}\label{sec:participants-conditions}

We recruited 76 English-speaking adults (aged 18 years or older) from the United States to participate in this online study. Sixty participants were recruited through Prolific and paid \$6 for a 30-minute task. The remaining 16 participants were drawn from a pool of undergraduate students and compensated with course credit. This research complied with the American Psychological Association Code of Ethics and was approved by the Institutional Review Board of Arizona State University. Informed consent was obtained digitally from all participants by requiring them to check a box before continuing with the study.

We reduced the number of conditions to a $3 \times 2$ design ($algorithms \times target~audiences$). Participants were randomly distributed across three conditions that implemented the baseline, ICL, and RP algorithms, and assigned to one of two target audiences. The two target audiences for this experiment, ``kids watching TV'' and ``busy managers'', were taken from the ones in the simulation and slightly adapted to include the most salient aspects of the omitted ones. Appendix~\ref{app:b} provides their descriptions.

Of the recruited participants, 64 completed the study. We excluded 18 people who completed the task unreasonably quickly (<5 minutes). This would not have been enough time to read the instructions and all 10 AI summaries, even if they were already in an appropriate style. An additional 16 participants were excluded for not making any changes to the first summary, which we took as an indicator that they did not take the task seriously. We assume that the first summary (for which there are no inferred preferences yet) always required edits. To minimize concerns that this assumption compromises the experiment, we provide the results of less conservative filtering in Appendix \ref{app:j}, showing that conclusions are not significantly impacted.

The final sample size was 30 participants. These exclusions disproportionally affected the RP conditions, especially for the target audience of children. This might be due to longer loading times that participants experienced because this algorithm requires an additional inference step. In our recruitment, we aimed to even out the distribution of participants across groups, which resulted in the three group sizes ranging from 8 to 13. On average, participants took 28.24 minutes ($SD = 19.59$) to complete the study.

\subsection{Results}\label{sec:results2}

A repeated-measures correlation analysis \citep{vallat_pingouin_2018} showed that participants' appropriateness ratings of the original summaries correlated with BERTScore, r(269) = 0.37, p < .001, and Levenshtein distance, $r(269) = -0.32$, $p < .001$. The low coefficient indicates a discrepancy between the computational metrics and subjective ratings. This discrepancy raises the question to what extent BERTScore and Levenshtein distance capture the complex concept of perceived appropriateness. For this reason, and because there is no comparable measurement in the first experiment, we will leave this here as a tangential insight but not leverage it for further analysis.

\begin{figure}[t]

\centering

\includegraphics[width=0.95\linewidth]{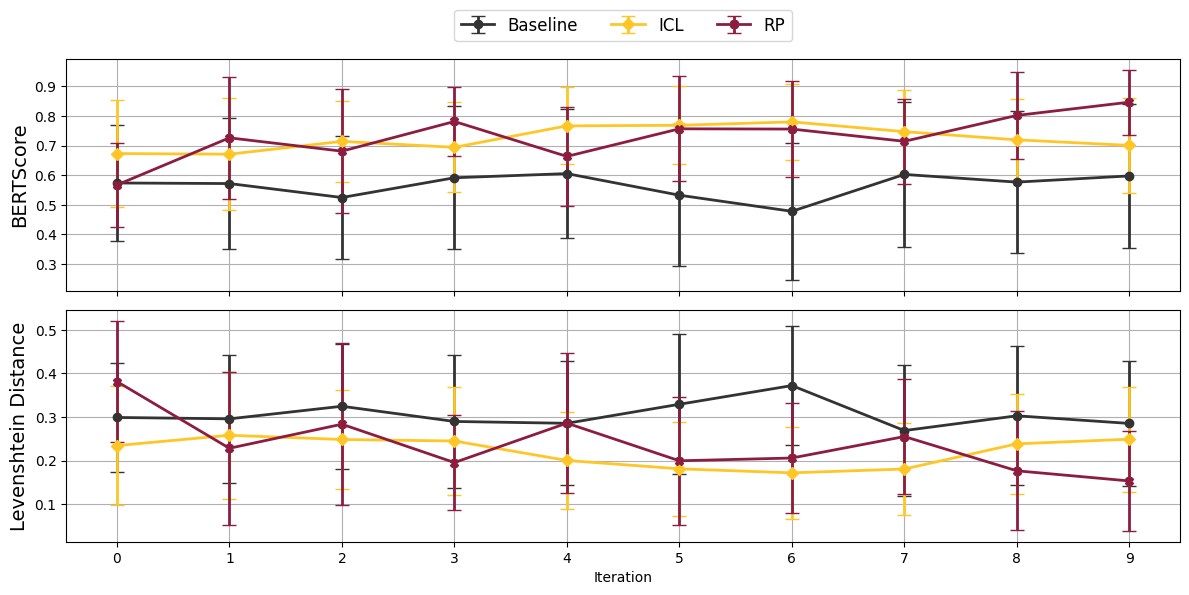}

\caption{Development of summary evaluation metrics over repeated interactions in the human trials. \textit{Similarity scores were calculated by comparing the summary originally generated by the algorithms to the edited version generated by the human participant. Error bars indicate 95\% confidence intervals. ICL = in-context learning, RP = rolling preferences.}
}

\label{fig:human-metrics}

\end{figure}

 BERTScore and Levenshtein distance exhibited a correlation almost as\linebreak strong as in the simulation, $r(269) = -0.92$, $p < .001$, so we report only the former for brevity. A mixed-effects model revealed that, across all conditions and iterations, participants judged the original summaries to be more appropriate for managers than for children, $b = -22.51$, $p = .001$. This is also reflected in the BERTScores, $b = -0.224$, $p = .011$, indicating that participants perceived and acted on a greater need to edit the summaries that were intended for children.

Figure~\ref{fig:human-metrics} shows the development of summary quality over repeated interactions. When comparing the mean BERTScore of the first iteration to the mean of the last interaction, a one-sided paired t-test clearly showed no significant improvement for the baseline, $t(8) = -0.57$, $p = .291$, or ICL, $t(12) = -0.51$, $p = .311$. Only the BERTScores in the RP condition improved over repeated interactions, $t(7) = -3.97$, $p = .003$. However, we are hesitant to conclude that this algorithm is superior because a qualitative inspection of participants' responses revealed signs of fatigue and questionable quality. All participants in this condition appeared to have used a strategy of simply cutting out sentences with little to no rephrasing. 

\begin{figure}[t]

\centering

\includegraphics[width=0.95\linewidth]{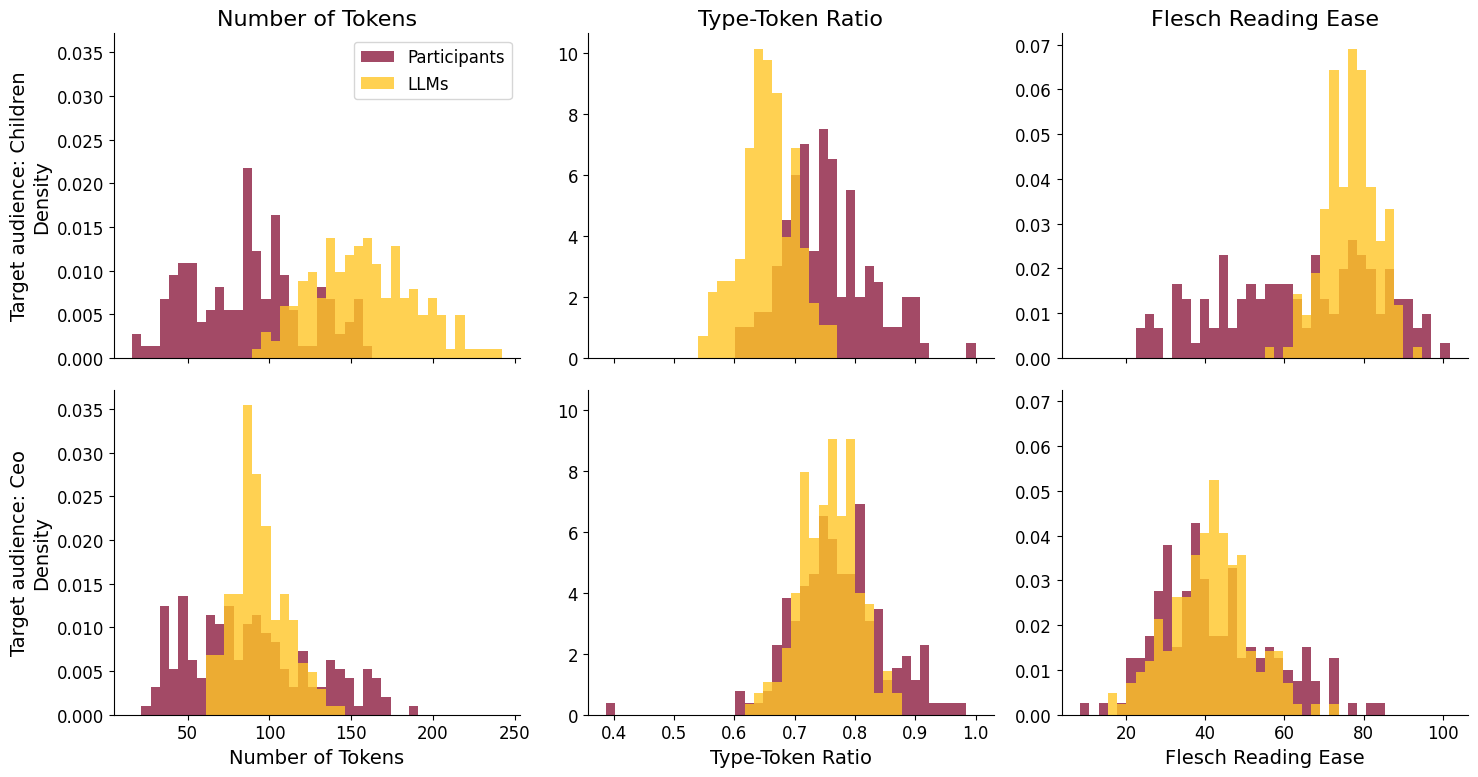}

\caption{Quantitative writing style analysis results.}

\label{fig:writing-style}

\end{figure}

An investigation of individual original-edit pairs with particularly low\linebreak BERTScores during the final iteration revealed that GPT-4o did not manage to capture the informal style that participants deemed appropriate for children. An illustrative example is provided in Appendix~\ref{app:f}. A qualitative exploration of inferred preferences (see Appendix~\ref{app:g}) hints at several other shortcomings. For example, the LLM correctly infers that ``the user shortened the summary significantly.'' However, without reference to the lengths of the original and final summaries, this did not lead to a reduction in the length of future summaries. These inferred preferences also include very little information or speculation on which parts the individual has removed and why.

A quantitative analysis of the summaries for a target audience of children edited by people ($n = 130$) and by GPT-4o (n = 180) hints at further fundamental stylistic differences between simulated and human participants (Figure~\ref{fig:writing-style}) which we tested with Mann-Whitney U-tests. Texts revised by simulated participants were longer (average length of 157.3 tokens compared to 87.6, $U = 1594.5$, $p < .001$), and had a lower lexical diversity (type-token ratio of .66 compared to .75, $U = 20730.5$, $p < .001$) as well as higher readability (Flesch reading ease score of 76.8 compared to 63.6, $U = 7149.0$, $p < .001$). These differences are much smaller for the target audience of CEOs but still statistically significant in the case of length ($U = 12782.5$, $p = .008$). These provide further potential explanations for why the LLM-powered style inference algorithms could not adapt as much to human participants' as to simulated participants' styles.

\section{Discussion}\label{sec:discussion}

\begin{figure}[t]

\centering

\includegraphics[width=0.95\linewidth]{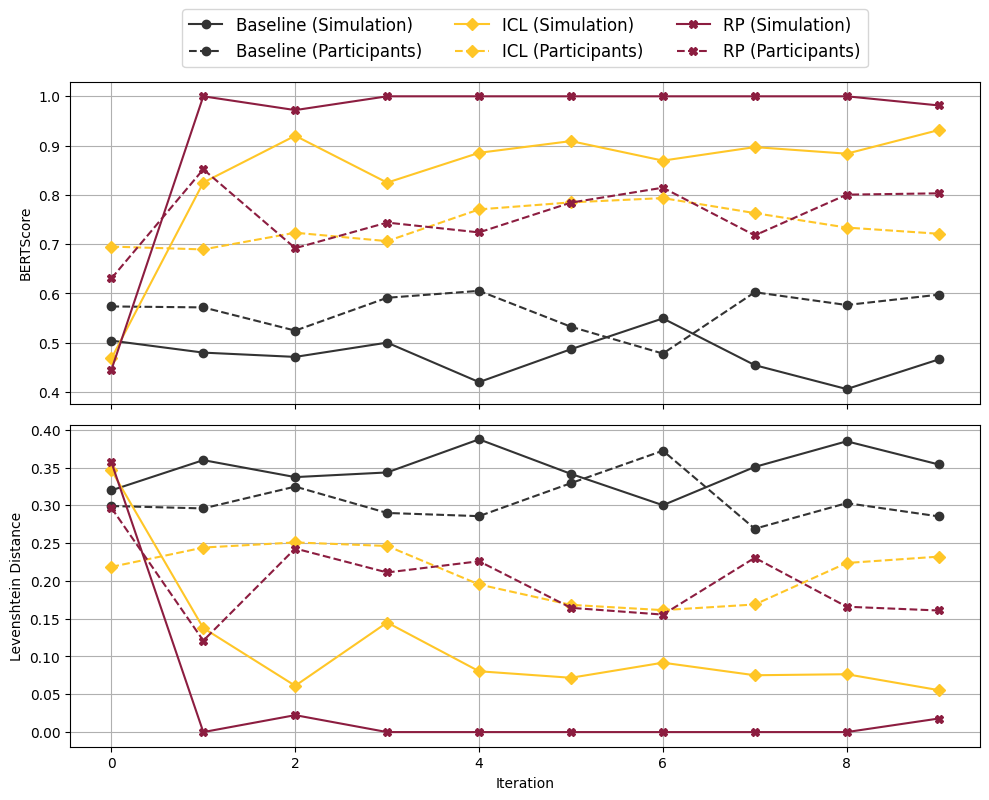}

\caption{Comparison of algorithm performance between the simulation and human trials. \textit{Similarity scores were calculated by comparing the summary originally generated by the algorithms to the edited version generated by the human participant. Error bars indicate 95\% confidence intervals. ICL = in-context learning, RP = rolling preferences.}
}

\label{fig:comparison}

\end{figure}

Figure~\ref{fig:comparison} combines the data of Figures~\ref{fig:sim-metrics} and~\ref{fig:human-metrics}, highlighting the divergence between the results of the simulation and experiment with human participants. This is in line with prior findings in related work \citep[e.g.,][]{aher_using_2023,cui_can_2025,manning_automated_2024,petrov_limited_2024} that simulations can be used to predict a rough trend (e.g., preference inference algorithms outperforming the baseline) but fail to predict the strength and small but important details (e.g., the development of ICL over time). Because we saw no significant improvement with ICL and are not confident in the findings for RP, our conclusion from the experiment with human participants is that the preference inference algorithms did not provide any benefit and are thus not worth the additional computational overhead. This is in direct contradiction to the findings of the simulation.

From our quantitative and qualitative findings, we conclude that our implementation of the simulated participant is poorly suited to realistically simulate people's editing behavior. The writing styles generated by GPT-4o are at the same time more extreme and more consistent than what people exhibit. The tireless nature of LLMs also makes them likely to put more effort into their revisions than people do. This exaggerated behavior makes it easier to infer the stylistic preferences of simulated users. Consequently, GPT-4o only needed a single contrastive example to adapt its writing style to a point where no further improvements could be made over the remaining 19 iterations. This indicates the threat that simulated users are likely to produce exaggerated results.

The LLM-based synthetic participants as well as the preference inference algorithms missed the detail, depth, and motivation of people's preferences. GPT-4o particularly failed to infer and replicate human imperfections that one naturally integrates when talking to certain audiences (e.g., children). The rolling preferences algorithm produced mostly descriptive style preferences that often missed the motivation behind the changes as well as the context to anchor them. We also found that a common revision by human participants is to merely cut out parts without rephrasing anything. This supports \citet{aroca-ouellette_predict_2024}’s argument that being presented with a pre-generated summary induces complacency and a satisficing behavior in which participants make minimal revisions that satisfy the requirements \citep{moray_monitoring_2003}.

GPT-4o, on the other hand, seemed overeager to revise, rephrasing even if the original summary was already appropriate for the target audience. This behavior was corrected through the two-stage design of the simulated user \citep{gao_aligning_2024}, indicating that smart design choices can improve the realism of synthetic participants. It is unlikely that off-the-shelf LLMs accurately simulate human nature. More research needs to be done to identify engineering approaches that replicate these important nuances we observe in human participants. At the same time, researchers must be careful to avoid introducing biases when engineering their simulated users to be more complex in an attempt to create more realistic simulations. 

Diverging results open the question whether the simulation or the experiment with human participants produced more useful results for real-world applications. We were concerned about participant fatigue but found that this did not lead participants in the baseline condition to make fewer edits towards the end of the 30-minute experiment. In the rolling preferences condition, however, our implementation might have caused a worse user experience, leading to increased fatigue and a survivorship bias in favor of low-effort participants who were not too invested in the task in the first place. This is a plausible explanation for the significant improvement in the rolling preferences condition that we found. This insight highlights the importance of considering human factors in the evaluations of algorithms. An algorithm that performs well in simulations might not be preferred by people in real-world use-cases if it lacks usability because it is cumbersome to use. Researchers also need to consider human factors for the choice of evaluation metrics. In our experiment, no changes made to an original summary could indicate either a high-quality summary or a low-effort revision. 

However, our method shows how simulations can be used as a time- and cost-effective method to compare and select algorithms or prompt variations. We were able to show significant benefits of some algorithms over the baseline and compared different algorithms and target audience descriptions. This helped us prepare for the experiment with human participants. 

\section{Limitations}\label{sec:limitations}

One clear limitation of our study is its small scope. We recruited a relatively small sample and spread participants across three different conditions. We did not force participants to make edits to keep the task setting more realistic and explicitly instructed them to ``not take too long for each summary'' to prevent them from exceeding the expected 30 minutes for the experiment. However, this also resulted in a large share of our participants applying minimal to no changes, reflecting a natural tendency \citep{moray_monitoring_2003} that nonetheless affects the interpretation of our results.

We acknowledge that our simulation is limited to a single LLM and that other models, especially those from other providers, might be more capable of capturing and applying stylistic details. Our findings are also limited to the domain of style preference inference over repeated interactions. With the collected data we cannot claim whether similar problems exist for LLM simulations in other domains, although we believe this to be highly likely. We believe that the research issue of LLM simulation fidelity benefits from insights in a broad variety of task domains.

We also reduced the number of tasks, scenarios, and target audiences of the PRELUDE framework, and omitted the mechanism that retrieves similar examples \citep{gao_aligning_2024}. We did this to demonstrate the discrepancy between simulated and real people with limited resources. Our findings regarding the performance of specific algorithms is thus not applicable. These simplified versions should be expected to perform worse which might have led to a deteriorated user experience. We attribute the significant findings of the rolling preferences condition mostly to such uncontrolled aspects of our experiment. This and the high exclusion rate threaten the validity of our experiment with human participants. However, we are confident that our main argument still holds because we show a clear discrepancy between the findings of the two experiments. This raises the question whether the simulation or the experiment with human participants is closer to the truth and provides more valuable insights. Biases could be introduced in both, the experimental setting as well as the simulation design. This further emphasizes the need for researchers to make their experimental and simulation designs transparent for the research community to discuss and criticize.

\section{Conclusion}\label{sec:conclusion}

In this study, we used the PRELUDE framework \citep{gao_aligning_2024} to evaluate different writing style preference inference algorithms. We compared the results of a simulation in which GPT-4o was tasked to emulate human behavior to the results of a replication experiment with human participants. Our comparison revealed a large discrepancy between the results of the two experiments. The simulation revealed biases and other aspects of LLMs that skew the performance evaluations, highlighting the threat to ecological validity if conclusions are solely based on the behavior of simulated users. 

The replication with real people, on the other hand, demonstrated how aspects of human nature, such as satisficing behavior, intent of language, and usability concerns, threaten the validity of findings derived purely from simulations. Researchers need to account for the human factors when studying human-LLM interaction. Relevant biases of people and LLMs might only be revealed in studies where they actually interact. We suspect that our experimental design and setup further exacerbated the divergence of behavior and thus caution future work to carefully consider and discuss the validity of synthetic participants wherever they are used as a research method.

\bibliographystyle{apalike}

\bibliography{references}

\begin{thebibliography}{}

\bibitem[Aher et~al., 2023]{aher_using_2023}
Aher, G.~V., Arriaga, R.~I., and Kalai, A.~T. (2023).
\newblock Using large language models to simulate multiple humans and replicate human subject studies.
\newblock In {\em Proceedings of the 40th {International} {Conference} on {Machine} {Learning}}, pages 337--371. PMLR.

\bibitem[Anthis et~al., 2025]{anthis_llm_2025}
Anthis, J.~R., Liu, R., Richardson, S.~M., Kozlowski, A.~C., Koch, B., Evans, J., Brynjolfsson, E., and Bernstein, M. (2025).
\newblock {LLM} social simulations are a promising research method.
\newblock arXiv:2504.02234.

\bibitem[Argyle et~al., 2023]{argyle_out_2023}
Argyle, L.~P., Busby, E.~C., Fulda, N., Gubler, J., Rytting, C., and Wingate, D. (2023).
\newblock Out of one, many: {Using} language models to simulate human samples.
\newblock {\em Political Analysis}, 31(3):337--351.
\newblock arXiv:2209.06899.

\bibitem[Aroca-Ouellette et~al., 2024]{aroca-ouellette_predict_2024}
Aroca-Ouellette, S., Mackraz, N., Theobald, B.-J., and Metcalf, K. (2024).
\newblock {PREDICT}: {Preference} reasoning by evaluating decomposed preferences inferred from candidate trajectories.
\newblock http://arxiv.org/abs/2410.06273.

\bibitem[Brown et~al., 2020]{brown_language_2020}
Brown, T., Mann, B., Ryder, N., Subbiah, M., Kaplan, J.~D., Dhariwal, P., Neelakantan, A., Shyam, P., Sastry, G., Askell, A., Agarwal, S., Herbert-Voss, A., Krueger, G., Henighan, T., Child, R., Ramesh, A., Ziegler, D., Wu, J., Winter, C., Hesse, C., Chen, M., Sigler, E., Litwin, M., Gray, S., Chess, B., Clark, J., Berner, C., McCandlish, S., Radford, A., Sutskever, I., and Amodei, D. (2020).
\newblock Language models are few-shot learners.
\newblock In {\em Advances in {Neural} {Information} {Processing} {Systems}}, volume~33, pages 1877--1901.

\bibitem[Cheng et~al., 2023]{cheng_compost_2023}
Cheng, M., Piccardi, T., and Yang, D. (2023).
\newblock {CoMPosT}: {Characterizing} and evaluating caricature in {LLM} simulations.
\newblock {\em Proceedings of the 2023 Conference on Empirical Methods in Natural Language Processing}, pages 10853--10875.

\bibitem[Cui et~al., 2025]{cui_can_2025}
Cui, Z., Li, N., and Zhou, H. (2025).
\newblock Can large language models replace human subjects? {A} large-scale replication of scenario-based experiments in psychology and management.
\newblock arXiv:2409.00128.

\bibitem[Dillion et~al., 2023]{dillion_can_2023}
Dillion, D., Tandon, N., Gu, Y., and Gray, K. (2023).
\newblock Can {AI} language models replace human participants?
\newblock {\em Trends in Cognitive Sciences}, 27(7):597--600.

\bibitem[Dong et~al., 2024]{dong_can_2024}
Dong, Y.~R., Hu, T., and Collier, N. (2024).
\newblock Can {LLM} be a {Personalized} {Judge}?
\newblock arXiv:2406.11657.

\bibitem[Gao et~al., 2024]{gao_aligning_2024}
Gao, G., Taymanov, A., Salinas, E., Mineiro, P., and Misra, D. (2024).
\newblock Aligning {LLM} agents by learning latent preference from user edits.
\newblock In {\em Advances in {Neural} {Information} {Processing} {Systems}}, volume~37, pages 136873--136896.

\bibitem[Horton, 2023]{horton_large_2023}
Horton, J.~J. (2023).
\newblock Large language models as simulated economic agents: {What} can we learn from homo silicus?
\newblock arXiv:2301.07543 [econ].

\bibitem[Jiang et~al., 2024]{jiang_survey_2024}
Jiang, R., Chen, K., Bai, X., He, Z., Li, J., Yang, M., Zhao, T., Nie, L., and Zhang, M. (2024).
\newblock A survey on human preference learning for large language models.
\newblock arXiv:2406.11191.

\bibitem[Levenshtein, 1966]{levenshtein_binary_1966}
Levenshtein, V.~I. (1966).
\newblock Binary codes capable of correcting deletions, insertions, and reversals.
\newblock {\em Soviet physics doklady}, 10(8):707--710.

\bibitem[Lin et~al., 2023]{lin_unlocking_2023}
Lin, B.~Y., Ravichander, A., Lu, X., Dziri, N., Sclar, M., Chandu, K., Bhagavatula, C., and Choi, Y. (2023).
\newblock The {Unlocking} {Spell} on {Base} {LLMs}: {Rethinking} {Alignment} via {In}-{Context} {Learning}.
\newblock arXiv:2312.01552.

\bibitem[Liu et~al., 2025]{liu_survey_2025}
Liu, J., Qiu, Z., Li, Z., Dai, Q., Zhu, J., Hu, M., Yang, M., and King, I. (2025).
\newblock A survey of personalized large language models: progress and future directions.
\newblock arXiv:2502.11528.

\bibitem[Lu et~al., 2024]{lu_llms_2024}
Lu, Y., Aleta, A., Du, C., Shi, L., and Moreno, Y. (2024).
\newblock {LLMs} and generative agent-based models for complex systems research.
\newblock {\em Physics of Life Reviews}, 51:283--293.

\bibitem[Manning et~al., 2024]{manning_automated_2024}
Manning, B.~S., Zhu, K., and Horton, J.~J. (2024).
\newblock Automated social science: language models as scientists and subjects.
\newblock Technical Report 32381, National Bureau of Economic Research, Cambridge, MA.

\bibitem[Moray, 2003]{moray_monitoring_2003}
Moray, N. (2003).
\newblock Monitoring, complacency, scepticism and eutactic behaviour.
\newblock {\em International Journal of Industrial Ergonomics}, 31(3):175--178.

\bibitem[{OpenAI}, 2024]{openai_hello_2024}
{OpenAI} (2024).
\newblock Hello {GPT}-4o.
\newblock https://openai.com/index/hello-gpt-4o/.

\bibitem[Petrov et~al., 2024]{petrov_limited_2024}
Petrov, N.~B., Serapio-García, G., and Rentfrow, J. (2024).
\newblock Limited ability of {LLMs} to simulate human psychological behaviours: {A} psychometric analysis.
\newblock arXiv:2405.07248.

\bibitem[Salemi et~al., 2024]{salemi_lamp_2024}
Salemi, A., Mysore, S., Bendersky, M., and Zamani, H. (2024).
\newblock {LaMP}: {When} large language models meet personalization.
\newblock {\em Proceedings of the 62nd Annual Meeting of the Association for Computational Linguistics}, Volume 1: Long Papers:7370--7392.

\bibitem[Shen et~al., 2024]{shen_pmg_2024}
Shen, X., Zhang, R., Zhao, X., Zhu, J., and Xiao, X. (2024).
\newblock {PMG}: {Personalized} multimodal generation with large language models.
\newblock {\em Proceedings of the ACM Web Conference 2024}, pages 3833--3843.

\bibitem[Stiennon et~al., 2020]{stiennon_learning_2020}
Stiennon, N., Ouyang, L., Wu, J., Ziegler, D., Lowe, R., Voss, C., Radford, A., Amodei, D., and Christiano, P.~F. (2020).
\newblock Learning to summarize with human feedback.
\newblock In {\em Advances in {Neural} {Information} {Processing} {Systems}}, volume~33, pages 3008--3021.

\bibitem[Tan et~al., 2025]{tan_democratizing_2025}
Tan, Z., Zeng, Q., Tian, Y., Liu, Z., Yin, B., and Jiang, M. (2025).
\newblock Democratizing large language models via personalized parameter-efficient fine-tuning.
\newblock {\em Proceedings of the 2024 Conference on Empirical Methods in Natural Language Processing}, pages 6476--6491.
\newblock arXiv:2402.04401.

\bibitem[Vallat, 2018]{vallat_pingouin_2018}
Vallat, R. (2018).
\newblock Pingouin: {Statistics} in {Python}.
\newblock {\em Journal of Open Source Software}, 3(31):1026.

\bibitem[Wang et~al., 2025a]{wang_user_2025}
Wang, L., Zhang, J., Yang, H., Chen, Z.-Y., Tang, J., Zhang, Z., Chen, X., Lin, Y., Sun, H., Song, R., Zhao, X., Xu, J., Dou, Z., Wang, J., and Wen, J.-R. (2025a).
\newblock User behavior simulation with large language model-based agents.
\newblock {\em ACM Transactions on Information Systems}, 43(2):1--37.

\bibitem[Wang et~al., 2025b]{wang_what_2025}
Wang, Q., Wu, J., Tang, Z., Luo, B., Chen, N., Chen, W., and He, B. (2025b).
\newblock What limits {LLM}-based human simulation: {LLMs} or our design?
\newblock arXiv:2501.08579.

\bibitem[Xie et~al., 2024]{xie_can_2024}
Xie, C., Chen, C., Jia, F., Ye, Z., Lai, S., Shu, K., Gu, J., Bibi, A., Hu, Z., Jurgens, D., Evans, J., Torr, P., Ghanem, B., and Li, G. (2024).
\newblock Can large language model agents simulate human trust behavior?
\newblock In {\em Advances in {Neural} {Information} {Processing} {Systems}}, volume~37, pages 15674--15729.

\bibitem[Yang et~al., 2024]{yang_llm_2024}
Yang, J.~C., Dailisan, D., Korecki, M., Hausladen, C.~I., and Helbing, D. (2024).
\newblock {LLM} voting: {Human} choices and {AI} collective decision making.
\newblock {\em Proceedings of the AAAI/ACM Conference on AI, Ethics, and Society}, 7:1696--1708.

\bibitem[Ye et~al., 2024]{ye_justice_2024}
Ye, J., Wang, Y., Huang, Y., Chen, D., Zhang, Q., Moniz, N., Gao, T., Geyer, W., Huang, C., Chen, P.-Y., Chawla, N.~V., and Zhang, X. (2024).
\newblock Justice or prejudice? {Quantifying} biases in {LLM}-as-a-judge.
\newblock arXiv:2410.02736.

\bibitem[Zhang et~al., 2020]{zhang_bertscore_2020}
Zhang, T., Kishore, V., Wu, F., Weinberger, K.~Q., and Artzi, Y. (2020).
\newblock {BERTScore}: {Evaluating} text generation with {BERT}.
\newblock arXiv:1904.09675.

\bibitem[Zhang et~al., 2024]{zhang_personalization_2024}
Zhang, Z., Rossi, R.~A., Kveton, B., Shao, Y., Yang, D., Zamani, H., Dernoncourt, F., Barrow, J., Yu, T., Kim, S., Zhang, R., Gu, J., Derr, T., Chen, H., Wu, J., Chen, X., Wang, Z., Mitra, S., Lipka, N., Ahmed, N., and Wang, Y. (2024).
\newblock Personalization of large language models: {A} survey.
\newblock arXiv:2411.00027.

\bibitem[Zhao et~al., 2025]{zhao_llms_2025}
Zhao, S., Hong, M., Liu, Y., Hazarika, D., and Lin, K. (2025).
\newblock Do {LLMs} recognize your preferences? {Evaluating} personalized preference following in {LLMs}.
\newblock arXiv:2502.09597.

\bibitem[Zheng et~al., 2023]{zheng_judging_2023}
Zheng, L., Chiang, W.-L., Sheng, Y., Zhuang, S., Wu, Z., Zhuang, Y., Lin, Z., Li, Z., Li, D., Xing, E., Zhang, H., Gonzalez, J.~E., and Stoica, I. (2023).
\newblock Judging {LLM}-as-a-judge with {MT}-{Bench} and {Chatbot} {Arena}.
\newblock {\em Advances in Neural Information Processing Systems}, 36:46595--46623.

\bibitem[Zhuang et~al., 2024]{zhuang_hydra_2024}
Zhuang, Y., Sun, H., Yu, Y., Qiang, R., Wang, Q., Zhang, C., and Dai, B. (2024).
\newblock {HYDRA}: {Model} factorization framework for black-box {LLM} personalization.
\newblock In {\em Advances in {Neural} {Information} {Processing} {Systems}}, volume~37, pages 100783--100815.

\end{thebibliography}

\clearpage

\appendix

\clearpage\section{Wikipedia articles used for the experiments}\label{app:a}

The following are the Wikipedia articles that have been used in our experiments. The articles were all retrieved in March 2025. The editing process consisted of shortening, removing formatting, and removing footnotes. The first 10 articles in this list were used in the experiment with human participants. 

\begin{table}[h]
\centering
\label{tab:wikipedia-articles}
\caption{Wikipedia articles used in the study.}
\small
\begin{tabularx}{\linewidth}{>{\raggedright\arraybackslash}p{0.32\linewidth}X}
\hline
\textbf{Topic} & \textbf{URL} \\
\hline
300-page iPhone bill &
\url{https://en.wikipedia.org/wiki/300-page_iPhone_bill} \\

NotebookLM &
\url{https://en.wikipedia.org/wiki/NotebookLM} \\

Grimace Shake &
\url{https://en.wikipedia.org/wiki/Grimace_Shake} \\

Hans Zimmer &
\url{https://en.wikipedia.org/wiki/Hans_Zimmer} \\

5D Chess Game &
\url{https://en.wikipedia.org/wiki/5D_Chess_with_Multiverse_Time_Travel} \\

The .io internet domain &
\url{https://en.wikipedia.org/wiki/.io} \\

Emoji-only social network &
\url{https://en.wikipedia.org/wiki/Emojli} \\

``I Am Rich'' app &
\url{https://en.wikipedia.org/wiki/I_Am_Rich} \\

Apple Vision Pro &
\url{https://en.wikipedia.org/wiki/Apple_Vision_Pro} \\

OpenAI o1 &
\url{https://en.wikipedia.org/wiki/OpenAI_o1} \\

Shortest regular ferry &
\url{https://en.wikipedia.org/wiki/Ferry_Lina} \\

Microsoft's Chatbot Tay &
\url{https://en.wikipedia.org/wiki/Tay_(chatbot)} \\

Nikon camera lens mount &
\url{https://en.wikipedia.org/wiki/Nikon_Z-mount} \\

iPhone 16 &
\url{https://en.wikipedia.org/wiki/IPhone_16} \\

Jeep Wagoneer S &
\url{https://en.wikipedia.org/wiki/Jeep_Wagoneer_S} \\

Epic Mickey Game &
\url{https://en.wikipedia.org/wiki/Epic_Mickey\#Remake} \\

Balatro Game &
\url{https://en.wikipedia.org/wiki/Balatro} \\

Hot Wheels &
\url{https://en.wikipedia.org/wiki/Hot_Wheels} \\

Moo Box &
\url{https://en.wikipedia.org/wiki/Moo_box} \\

Breakout video game &
\url{https://en.wikipedia.org/wiki/Breakout_(video_game)} \\
\hline
\end{tabularx}
\end{table}

\clearpage\section{Descriptions of target audiences}\label{app:b}

The following descriptions were given to GPT-4o in the simulation.

\begin{itemize}

\item \textbf{Young children watching TV:} Your summaries will be read out loud in an educational TV show. Make sure that the words you use are not too advanced or complicated. Your goal is to make these segments exciting and fun to listen to.

\item \textbf{High-school students in class:} Your content should be engaging and include at least one open question. Your goal is to make sure this summary sparks a discussion in class.

\item \textbf{Managers and CEOs:} Your bosses want to know what this is all about but they get hundreds of memos each day. Your goal is to make the reader understand all the facts in less than 30 seconds of reading.

\item \textbf{Young adults on social media:} Make your summary engaging by highlighting emotions and desires. Your goal is to get as many people as possible to leave a comment on your post

\end{itemize}

The following descriptions of target audience were given to human participants.

\begin{itemize}
\item \textbf{Young children watching TV:} Your summaries will be read out loud in an educational TV show. Make sure that the words you use are not too advanced or complicated. Your goal is to make these segments exciting and fun to listen to. Include at least one open question to spark a discussion!

\item \textbf{Managers and CEOs:} Your bosses want to know what this is all about but they get hundreds of memos each day. Your goal is to make the reader understand all the facts in less than 30 seconds of reading.

\end{itemize}

\clearpage\section{Prompt templates}\label{app:c}
\subsection{Baseline}
\begin{quote}
Article:

\texttt{\{article\_text\}}

Please summarize the above article.
\end{quote}

\subsection{In-context learning}
\begin{quote}
Your original summary of an article:

\texttt{\{original\_summary\}}

Revised summary by the user:

\texttt{\{edited\_summary\}}

\texttt{---}

\texttt{\{\ldots{} more pairs\}}

\texttt{---}

Based on the edits and revision by this user on the original summary in the above examples, please summarize the following article:

\texttt{\{article\_text\}}
\end{quote}

\subsection{Rolling preferences generation}
\begin{quote}
\texttt{\{baseline\_prompt\}}

From previous user edits you have inferred the following preferences. Closely follow and strongly emphasize them.

\texttt{\{rolling\_preferences\}}
\end{quote}

\subsection{Rolling preferences inference}
\begin{quote}
You generated the following summary of an article:

\texttt{\{original\_summary\}}

After the user revised it, the summary now looks like this:

\texttt{\{edited\_summary\}}

Identify the differences between the texts and infer the user's writing preferences? Consider diction, lexical density, sentence length and complexity, use of passive vs. active voice, imagery, figurative language, point of view, tense consistency, structural elements, formatting preferences, emotional tone, formality level, and punctuation.

Please answer only with a list of the user's preferences you have inferred from the changes they have made (or not made).
\end{quote}

\subsection{Rolling preferences consolidation}
\begin{quote}
We are tasked to curate a prompt to guide a specific style of writing. We currently have the following list of preferences related to writing styles:

\texttt{\{rolling\_preferences\}}

From the last interaction you have inferred the following Insights:

\texttt{\{inferred\_preferences\}}

Unfortunately, these preferences may overlap or contain redundancies. Please review the list and condense it by combining similar or overlapping preferences, ensuring that the distinct intent behind each one remains clear so that a writer can easily follow them. Ensure the condensed list is concise, non-redundant, and preserves the original level of specificity. When applicable, preserve the exact wording. Return the revised preferences in the same format as the original list.
\end{quote}

\subsection{Appropriateness check}
\begin{quote}
Summary:

\texttt{\{original\_summary\}}

Is the above summary in an appropriate style for the following target audience?

\texttt{\{audience\_description\}}

Please answer only yes or no.
\end{quote}

\subsection{Summary revision}
\begin{quote}
Original summary:

\texttt{\{original\_summary\}}

Please revise the above summary of an article to meet the requirements of the following target audience:

\texttt{\{audience\_description\}}
\end{quote}

\clearpage\section{Effects the two-stage simulated user implementation}\label{app:d}

\begin{figure}[bth]

\centering

\includegraphics[width=0.95\linewidth]{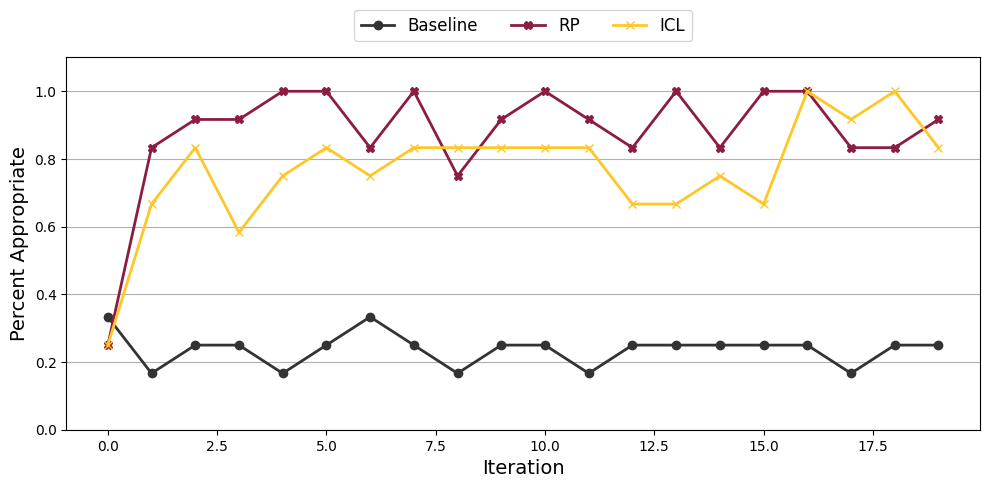}

\caption{Percentage of original summaries accepted by simulated users without revisions. \textit{ICL = in-context learning, RP = rolling preferences.}}

\label{fig:accepted}

\end{figure}

The two-stage implementation does a first inference pass merely asking whether the original summary is sufficiently appropriate for the target audience. If the synthetic participant responds with ``yes'' the original summary is copied without changes, resulting in a BERTScore of 1 and a Levenshtein distance of 0. Figure~\ref{fig:accepted} shows the percentage of summaries which were deemed sufficiently appropriate without the need for further revisions. 

\begin{figure}[thb]

\centering

\includegraphics[width=0.95\linewidth]{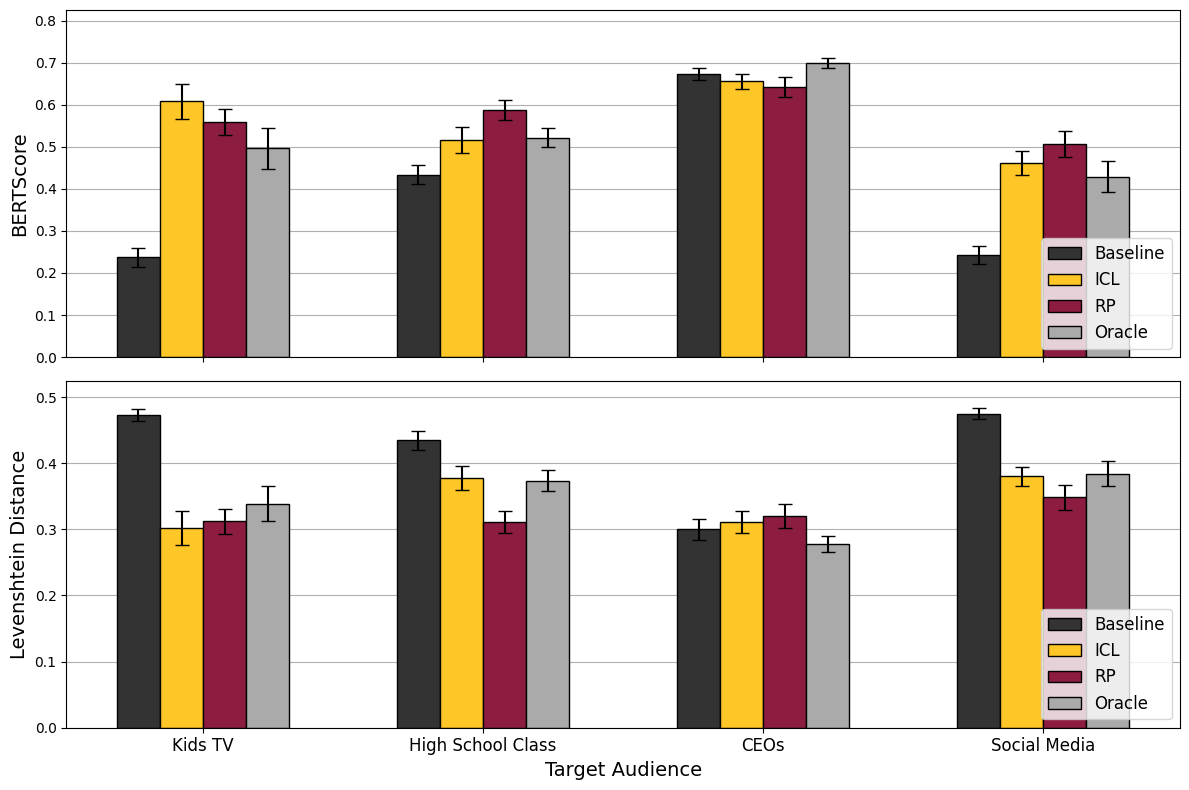}

\caption{Summary evaluations across all articles without appropriateness check stage. \textit{Similarity scores were calculated by comparing the summary originally generated by the algorithms to the edited version generated by the simulated participant, and averaged across all iterations. Error bars indicate 95\% confidence intervals. ICL = in-context learning, RP = rolling preferences.}}

\label{fig:no-check}

\end{figure}

Figure~\ref{fig:summary-evals} in the main body of the article shows the resulting performance per persona and inference algorithm using the two-stage implementation. Because this strategy inhibits the tendency for over-eager revisions, the effects are dampened when skipping the appropriateness check (Figure~\ref{fig:no-check}).

\clearpage\section{Example of an unnecessarily heavily edited\linebreak summary by a simulated user}\label{app:e}

The following is an example of a summary generated by the\linebreak rolling-preferences algorithm for the target audience of young children\linebreak watching TV, and then edited by the simulated user without the two-stage implementation (i.e., asking the synthetic participant to revise until appropriate without checking if it was already appropriate). We suggest that these revisions, which lead to an exceptionally low BERTScore, can likely be attributed to the temperature setting. We have removed 12 and 13 emojis respectively from each summary to ensure the compatibility of this paper.

\subsection{Original summary}

Hey there, awesome explorers of the digital world! Buckle up for a whirlwind tour of the .io domain---a name you might have seen all over your favorite games and tech sites. But did you know it's connected to a slice of oceanic paradise called the British Indian Ocean Territory? Sounds exotic, right?

Here's the scoop: The .io domain first popped up in internet land thanks to a dude named Paul Kane, who got the keys to not just .io, but to some other intriguing letter combos like .ac and .sh! Fast forward to 2014, Kane said, ``Hey, I'm helping the islands run smoothly by distributing some profits there!'' But wait for it... the UK government said, ``Nope, not true!'' Plot twist!

Now, just when you think the roller coaster is over, the UK plans to hand over this dreamy territory to Mauritius. But here's the kicker: If the digital rulekeepers have their way, our beloved .io might do a disappearing act in about 5 years. Poof! Even though there's a chance it could stick around like its quirky cousin .su did.

Oh, and our friends across the pond in the US are thinking this through after their big election shake-up, so who knows what the future holds?

So there you have it, tech explorers---dot your I's and cross your O's, let's keep our eyes peeled on .io's next adventure! Keep surfing the web waves, and have an awesome day!

\subsection{Revised summary by a simulated user}

Hello, little adventurers of the digital playground! Are you ready for a fun story about .io, a special part of the internet you might see in games? It's like a secret island treasure!

Once upon a time, a clever person named Paul found this cool .io island and a few other islands like .ac and .sh---because who doesn't love a good island adventure? For a while, people thought Paul was sharing some treasure with the real island, but the grown-ups in charge said, "Nope, that's not right!" Surprise!

And guess what? The UK, the big helper of the island, wants to give it to another place called Mauritius. But here's the twist: .io might disappear in a few years like a magical trick! But maybe, just maybe, it will stay around longer, like a sneaky friend!

And far away in the US, people are thinking hard about what happens next after a big important event. Who knows what they will decide?

So keep your eyes open and your curiosity shining, young explorers. .io's adventure isn't over yet! Keep having fun online, and make every day amazing!

\clearpage\section{Example of an original summary that fails to capture the imperfect naturalness}\label{app:f}

The following example was generated in the ICL condition of the experiment with human participants in the final iteration. It showcases GPT-4o's lack of natural imperfection that the participant deemed appropriate for an audience of children. We removed 13 emojis that the participant inserted to ensure the compatibility of this paper.

\subsection{Original summary}

Imagine an app where you could only use emoji to chat---that's Emojli! Created by YouTubers Tom Scott and Matt Gray, this quirky app was all about emoji usernames and emoji-only conversations. Originating as a bit of a joke, Emojli caught on fast, amassing 70,000 reserved usernames even before its launch on August 29, 2014. Despite attracting venture capital offers after media buzz, the app was shut down on July 30, 2015, due to the high costs and effort of keeping it running.

\subsection{Revised summary by a participant}

A while back, there was a silly and fun app called Emojli, where you could only communicate in emoji---no words allowed!

It was made by two funny YouTubers, Tom and Matt. They thought, "What if we made a whole app just for emoji?" And do you know what? A lot of people thought it was a good idea! 70,000 people had already picked emoji-only usernames before it even came out! 

Though the app gained tremendous popularity, it was very hard and expensive to maintain. So after a year, they had to say goodbye to Emojli. But it certainly put a smile on people's faces during its existence!

\clearpage\section{Example of inferred preferences}\label{app:g}

This is an example of an original summary, the revised version, and the inferred preferences from this contrast. This example was produced in the first iteration of the rolling preferences condition for a target audience of children.

\subsection{Original summary}

NotebookLM is an AI-driven research and note-taking tool by Google that aids users in document interaction, generating summaries, explanations, and audio overviews in a podcast-like format. Initially released for enterprises and paid users in December 2024 as NotebookLM Plus, it expanded to individual users via Google One AI Premium in February 2025. The tool processes various file types, including PDFs and Google Docs, and gained attention for summarizing complex documents. Though designed for researchers, it is also used by companies and students.

\subsection{Revised summary by a participant}

The Google tool "NotebookLM" is an AI-driven product that helps users explain documents, summarize documents, and audio overviews in a podcast-like format. It's somewhat new and processes different file types, including PDFs and Google Docs.

\subsection{Inferred preferences}

\begin{itemize}

\item Preference for brevity and conciseness in language.

\item Use of simpler diction and reduced lexical density.

\item Shorter sentence length and less complex sentence structures.

\item Reduction of passive voice; more emphasis on active voice.

\item Elimination of specific details, such as release dates and user categories.

\item Minimal use of technical jargon or specific terminology.

\item Preference for a straightforward and factual tone.

\item Reduced formality level in writing style.

\item Consistent use of present tense.

\item Lack of imagery or figurative language.

\item Simplified formatting with no added structural elements.

\end{itemize}

\clearpage\section{Instructions given to participants before the\linebreak first summary}\label{app:h}

\begin{quote}
The following pages will all follow the same process. You will see a document and will be asked to summarize it for a specific audience. Your starting point will be a summary generated by Artificial Intelligence (AI) that you can edit. Think about what writing style, choice of words, format, etc. is best suited for that audience.

You should not take too long for each summary because there are 10 summaries in total. Keep in mind that everything you submit will be recorded and analyzed by researchers and by AI. Do not enter any of your personal information in these summaries.

our target audience for the rest of the study is the following:

\texttt{\{target\_audience\}}

No need to memorize this; you will see that description again on every page. Now take a minute to think about and note down what writing style is appropriate to address this audience.
\end{quote}

\clearpage
\section{Screenshot of the interface provided to participants}\label{app:i}


\begin{figure}[h]

\centering

\includegraphics[width=0.66\linewidth]{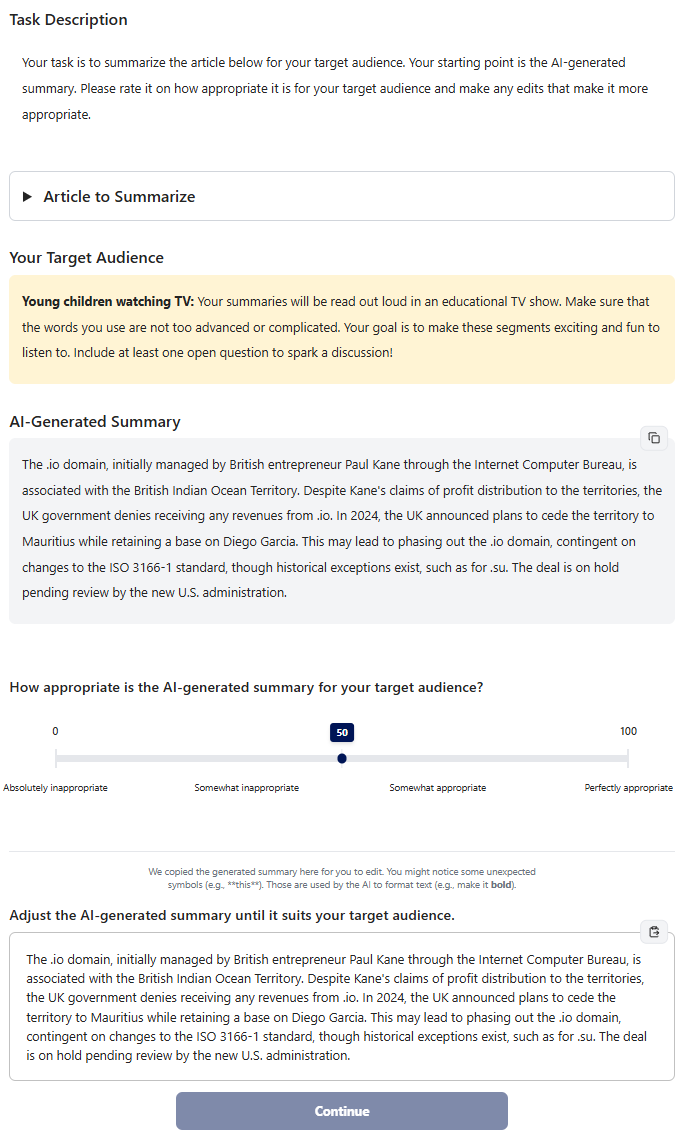}

\caption{Screenshot of an iteration page on our online survey platform.}

\label{fig:screenshot}

\end{figure}

\clearpage
\section{Overview of results without conservative participant exclusion}\label{app:j}

Because the exceptionally high percentage of excluded participants (keeping 30 of the 64 who completed the experiment) raises concerns about the validity of the experiment, we recalculated the results with less conservative requirements. In particular, we included the 16 participants who did not make changes to the first summary. The group sizes for the three algorithms are then 17, 15, and 14. Because the newly included data consists of mostly unedited final versions, all results become inflated. The correlation between BERTScore and participants' appropriateness ratings is considerably higher, $r(458) = 0.49$, $p < .001$, while the correlation between BERTScore and Levenshtein distance only received a small bump, $r(458) = -0.95$, $p < .001$. 

\begin{figure}[thb]

\centering

\includegraphics[width=0.95\linewidth]{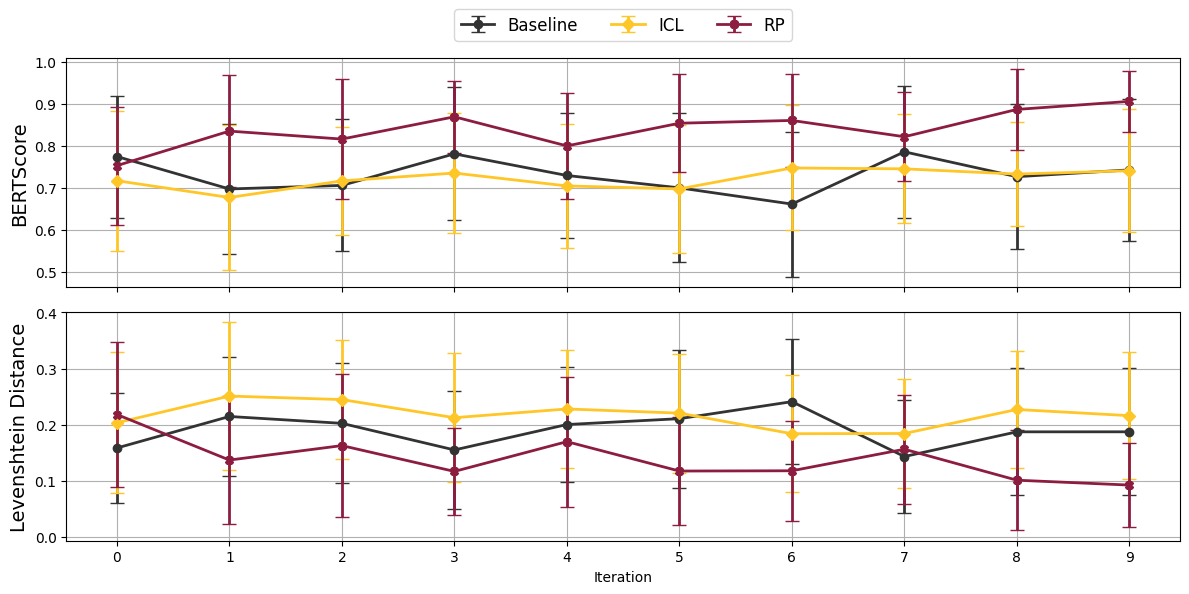}

\caption{Summary evaluation metrics without conservative exclusion criteria. \textit{Similarity scores were calculated by comparing the summary originally generated by the algorithms to the edited version generated by the human participant. Error bars indicate 95\% confidence intervals. ICL = in-context learning, RP = rolling preferences.}}

\label{fig:no-conservative}

\end{figure}

At every iteration, the average BERTScore of each performance is elevated, with the baseline and ICL conditions being affected more strongly (Figure~\ref{fig:no-conservative}). However, the relative performance of the three algorithms remains unchanged: only the RP condition shows significant improvement in BERTScore from the first to the last iteration, $t(14) = 2.24$, $p = .002$. Thus, the larger dataset would lead to the same conclusions as are reported in the results section of the paper.

\end{document}